# Inclination-Driven Thin-Film Hydrodynamics: Universal Trajectory in the {Da, Pe, Bo} Space

Helena Cristina Vasconcelos[1,2] *, Maria Meirelles[1,3] and Reşit Özmenteş[4]

[1]Faculty of Science and Technology, University of the Azores, Ponta Delgada, S. Miguel, 9500-321 Azores, Portugal
[2]Laboratory of Instrumentation, Biomedical Engineering and Radiation Physics (LIBPhys, UNL), Department of Physics, NOVA School of Science and Technology, 2829-516 Caparica, Portugal
[3]Research Institute of Marine Sciences of the University of the Azores (OKEANOS), Horta, Faial, 9901-862 Azores, Portugal
[4]Vocational School of Health Services, Bitlis Eren University, 13100 Bitlis, Turkey

**ABSTRACT**. We develop a unified theoretical framework for thin-film hydrodynamics on inclined solid substrates, integrating capillarity, intermolecular forces, gravitational symmetry breaking, confined transport and stochastic wetting into a single formulation. Starting from lubrication theory with capillary curvature and disjoining-pressure interactions, we obtain a general thin-film equation that incorporates inclination-driven advection, nanoscale stabilization and humidity-controlled source–sink fluxes. A dimensionless analysis shows that, within the long-wave lubrication approximation, inclination imposes a leading-order, geometry-induced coupling of the Bond, Péclet and Damköhler numbers. This coupling defines a characteristic trajectory in the {Da, Pe, Bo} space—set by the structure of the lubrication flux—along which material parameters determine the system's position but not the trajectory's underlying geometry. Coupling this deterministic structure to a minimal stochastic formulation captures the intermittent wet–dry dynamics characteristic of ultrathin films under environmental forcing. The resulting framework provides a general surface-physics description of confined films under geometric asymmetry, applicable to wetting, interfacial drainage, reactive confinement and soft-matter systems in which symmetry breaking and coupled interfacial–transport processes coexist across scales. The focus of this work is the development of the governing structure and the dimensionless framework, rather than the derivation of closed-form solutions or detailed numerical predictions. Quantitative solutions for specific regimes can be obtained from the same formulation and constitute natural extensions of the present conceptual framework.

## I. INTRODUCTION

Thin liquid films on solid surfaces constitute a canonical setting for studying the interplay between geometry, intermolecular forces and hydrodynamics. Their behavior governs wetting, adhesion, drainage, lubrication, interfacial transport and rupture, and has long served as a testbed for theories of capillarity, surface forces and thin-film flow [1-3]. When the film thickness reaches micro- and nanometric scales, the balance between capillary curvature, van der Waals interactions and viscous dissipation becomes highly sensitive to perturbations. Even weak geometric asymmetries can reorganize drainage pathways, stability criteria and rupture dynamics [4-6].

Among such asymmetries, inclination plays a distinctive physical role. Rather than acting as a minor geometric detail, inclination introduces a symmetry-breaking field that couples directly to the drainage flux through the tangential gravitational forcing $\rho g \sin\theta$. Classical analyses typically treat thin films in horizontal or weakly perturbed geometries [3,7,8], but natural and technological interfaces—ranging from atmospheric electrolyte layers to evaporating or deliquescent coatings—rarely remain horizontal. Such films undergo cycles of thinning, rupture, re-wetting and coalescence under the combined action of gravity, capillarity and disjoining pressure [7-10].

Experimental observations confirm that inclination strongly reorganizes thin-film behavior: inclined surfaces display faster thinning, earlier rupture and shorter wet-state persistence compared with horizontal substrates [11,12]. Similar inclination-dependent trends appear in lubrication systems [10], soft-matter and colloidal films [13], confined interfacial transport [14], evaporating and deliquescent layers, and thin-film flows governed by competing advection–diffusion mechanisms [15,16]. Despite this diversity, existing treatments remain fragmented: capillarity, disjoining pressure, gravitational drainage, confined electrostatics and stochastic wetting are often analyzed independently.

*Contact author: helena.cs.vasconcelos@uac.pt

Here we develop a unified theoretical framework that consolidates these mechanisms. Starting from lubrication theory with capillary curvature and disjoining-pressure interactions, we derive the thin-film governing equation in a form that makes the inclination-driven advection, nanoscale stabilization, and humidity-controlled source–sink fluxes explicit. A dimensionless formulation reveals that inclination induces a coordinated coupling of the Bond, Péclet and Damköhler numbers, driving the system along a universal trajectory in $\{\text{Bo}, \text{Pe}, \text{Da}\}$ space. This trajectory governs stability, drainage, reactive confinement and solute retention across material classes. Coupling this deterministic formulation to a minimal stochastic model captures the intermittent wet–dry cycling characteristic of real ultrathin films under environmental forcing.

The resulting formulation is general and material-independent, providing a surface-physics basis for understanding thin-film behavior under geometric asymmetry across hydrodynamics, interfacial physics, soft matter and confined transport systems.

The aim of this work is to establish a unified and physically transparent framework; detailed analytical solutions and numerical simulations fall outside the present scope but follow directly from the same governing formulation.

## II. GOVERNING EQUATIONS AND COUPLED FILM DYNAMICS

Thin-film dynamics on an inclined solid substrate arise from the coupled action of viscous flow, capillary curvature, intermolecular forces, gravitational symmetry breaking and humidity-controlled source–sink fluxes. This section derives the governing equations from lubrication theory and establishes the structure required for the dimensionless analysis in Sec. III.

### II.A. Lubrication framework and gravitational symmetry breaking

Consider a liquid film of local thickness $h(x,t)$ on a substrate inclined at an angle $\theta$. Under the long-wave approximation $h/L \ll 1$ and $|\partial_x h| \ll 1$, the lateral flow is well described by lubrication theory. A schematic of the inclined thin-film geometry and the relevant forces is shown in Fig. 1.

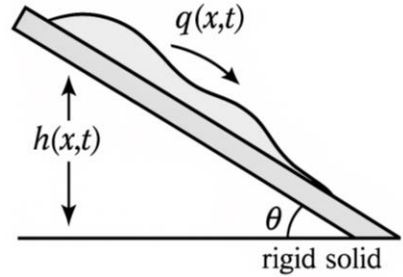

FIG 1. Schematic representation of a thin liquid film on an inclined solid substrate.

The depth-averaged volumetric flux per unit width, denoted $q(x,t)$, is [3,4]:
$$q(x,t) = -\frac{h^3}{3\mu}[\gamma\, \partial_{xxx}h - \partial_x \Pi(h) - \rho g \sin\theta] \quad (1)$$
where
- $\mu$ is viscosity,
- $\gamma$ is surface tension,
- $\Pi(h)$ is the disjoining pressure,
- $\rho g \sin\theta$ is the tangential gravitational forcing.

Mass conservation then gives:
$$\partial_t h + \partial_x q = S(x,t), \quad (2)$$
with $S(x,t)$ representing condensation–evaporation fluxes. Inclination enters solely through the $\sin\theta$ term, acting as a symmetry-breaking field that biases drainage without altering capillary or intermolecular contributions. This formulation applies broadly to ultrathin films dominated by viscous, capillary and van der Waals forces [1,2,5].

This structure also makes clear that inclination enters the governing flux only through the gravitational contribution, which becomes the leading-order source of θ-dependence in the dimensionless groups derived in Sec. III.

The thin-film flux in Eq. (1) is derived under the standard lubrication assumptions of Newtonian viscosity, no-slip at the solid surface, small interfacial slopes $|\partial_x h| \ll 1$, negligible inertia ($Re \ll 1$), and a separation of scales $h/L \ll 1$. Slip, if present, modifies only the mobility prefactor $h^3/3\mu$ and does not alter the leading-order structure of the capillary, disjoining-pressure or gravitational terms. The formulation is not intended to describe molecularly thin precursor layers or regimes in which continuum hydrodynamics, the disjoining-pressure model, or the classical mobility law break down. Within these constraints, Eq. (1) captures the dominant balance of forces governing micrometric and nanometric films on inclined substrates and provides the appropriate starting point for the nondimensional analysis of Sec. III.

*Contact author: helena.cs.vasconcelos@uac.pt

## II.B. Capillary curvature and intermolecular forces

Capillary pressure is
$$P_c = -\gamma\, \partial_{xx} h, \quad (3)$$
which stabilizes short-wavelength disturbances. Intermolecular forces enter through the disjoining pressure [2]:
$$\Pi(h) = -\frac{A}{6\pi h^3} + \Pi_{\text{rep}}(h), \quad (4)$$
with $A$ the Hamaker constant and $\Pi_{\text{rep}}(h)$ representing electrostatic or structural repulsion [6,17].

Absent gravity, the competition between $P_c$ and $\Pi(h)$ yields the classical spinodal-dewetting instability [7,9].

Inclination introduces a directional gravitational flux proportional to $\sin\theta$, reorganizing rupture pathways and stability thresholds [4,8,18].

## II.C. Source–sink terms and environmental coupling

Humidity-driven processes enter through
$$S = S_{\text{cond}}(RH, T) - S_{\text{evap}}(T, h), \quad (5)$$
describing condensation/deliquescence and evaporation [19,20].

These fluxes control re-wetting cycles, rupture recovery and wet/dry intermittency—behaviors characteristic of atmospheric electrolyte films [11,12].

## II.D. Confined electrostatics in ultrathin films

For electrolyte films, the electric potential $\phi(x,t)$ satisfies
$$\nabla \cdot [\kappa(h)\nabla\phi] = 0, \quad (6)$$
where $\kappa(h)$ is the thickness-dependent conductivity, reduced in nanometric films by electrical double-layer overlap [2].

Boundary conditions:
- solid–liquid interface: $-\kappa\, \partial_n \phi = j_{\text{int}}$
- liquid–air interface: $\partial_n \phi = 0$

Electric relaxation is much faster than hydrodynamic dynamics [14], justifying the quasi-static approximation.

## II.E. Ionic advection–diffusion and reactive transport

The solute concentration satisfies
$$\partial_t c + \partial_x(uc) = D\, \partial_{xx} c + R(c, h, \phi), \quad (7)$$
where
- $u = q/h$ is the depth-averaged lateral velocity,
- $D$ is the diffusivity,

*Contact author: helena.cs.vasconcelos@uac.pt

- $R$ represents reactive or electrochemical processes.

Inclination modifies solute retention through the ratio $uL/D$, linking gravitational drainage to mass-transport resistance and reactive confinement [15,16].

## II.F. Structure of the coupled thin-film fields

The coupled fields
$$\{h(x,t), c(x,t), \phi(x,t)\} \quad (8)$$
provide a complete continuum-scale description of the dynamics of inclined thin films. Within the lubrication framework, these equations capture the leading-order balance among viscous flow, capillarity, intermolecular forces, gravitational forcing and confined transport.

This formulation is valid under the usual long-wave assumptions: small interfacial slopes, negligible inertia, Newtonian viscosity and film thicknesses sufficiently large for continuum hydrodynamics to apply. Molecularly thin precursor layers, strong-slip conditions and Knudsen-scale effects lie outside the range of applicability of this leading-order model.

The nondimensional structure emerging from these equations is developed in Sec. III.

## III. DIMENSIONLESS FORMULATION AND UNIVERSAL INCLINATION TRAJECTORY

The dimensionless structure of the governing equations reveals how inclination reshapes the relative contributions of capillarity, intermolecular forces, gravitational drainage and confined transport. When expressed using appropriate characteristic scales, the thin-film equations reduce to a compact form governed by three natural dimensionless groups: the Bond, Péclet and Damköhler numbers.

The coordinated dependence of these groups on the inclination angle defines a leading-order trajectory in the {Bo, Pe, Da} parameter space. This geometry-imposed trajectory constrains the behavior of all inclined thin-film systems, setting the structure developed in the subsections below.

### III.A. Characteristic scales and nondimensional variables

A natural starting point for the nondimensional analysis is to identify the characteristic scales that describe the geometry and dynamics of the inclined film. Let $H$ denote a representative film thickness, $L$ a lateral length scale, $U$ the drainage velocity associated with gravitational forcing, and $T = L/U$ the corresponding advective timescale. Because of the gravitational contribution to the lubrication flux scales as $\rho g h^3 \sin\theta$, the velocity scale $U$ is chosen to preserve this leading-order dependence on $\theta$, ensuring that inclination appears explicitly—and solely—through gravity-driven advection in the dimensionless formulation.

Using these scales, the dimensional variables are expressed in nondimensional form as
$$x = Lx', h = Hh', t = Tt', q = UH\, q', \quad (9)$$
where primes are omitted after substitution.

With these definitions, the mass-conservation law becomes
$$\partial_t h + \partial_x q = \text{Da}_0, \quad (10)$$
where $\text{Da}_0$ is a nondimensional source–sink strength. The corresponding nondimensional flux takes the form
$$q = -h^3[\text{Ca}^{-1}\, \partial_{xxx}h - \partial_x \Pi^*(h) - \text{Bo}\sin\theta], \quad (11)$$
with
$$\text{Ca} = \frac{\mu U}{\gamma}, \text{Bo} = \frac{\rho g H L}{\gamma}, \Pi^*(h) = \frac{H}{\gamma}\Pi(h). \quad (12)$$
In this representation, inclination enters only through $\text{Bo}\sin\theta$, isolating the geometric contribution to the force balance independently of the specific material parameters.

### III.B. Balance between capillarity, intermolecular forces and gravity

The nondimensional flux highlights the competing roles of capillary curvature, intermolecular forces and gravitational drainage in determining thin-film dynamics.

In nondimensional form, these contributions scale as
- $\text{Ca}^{-1}\, \partial_{xxx}h$ (capillarity),
- $-\partial_x \Pi^*(h)$ (intermolecular forces),
- $-\text{Bo}\sin\theta$ (gravity),

while viscous resistance enters through the geometric factor $h^3$.

Because these terms enter additively into the flux, inclination modifies the relative magnitude of the competing contributions rather than their absolute scale. As $\theta$ increases, the gravitational term grows through $\text{Bo}\sin\theta$, while the capillary and intermolecular terms remain unchanged. This introduces a progressive reweighting of the force balance:
- the gravitational contribution strengthens monotonically,
- the effective stabilizing influence of capillarity weakens in relative terms,
- the film becomes increasingly sensitive to intermolecular attraction at small thickness.

The inclination angle $\theta$ therefore acts as a natural organising parameter for dynamics, altering the hierarchy of competing forces without affecting the structure of the governing equations.

### III.C. Solute Transport and the Péclet Number

Solute transport in the draining film is governed by the interplay between advection, diffusion and reaction. In nondimensional form, the transport equation reads
$$\partial_t c + \partial_x(uc) = \text{Pe}^{-1}\, \partial_{xx}c + R(c, h, \phi), \quad (13)$$
where the dimensionless velocity is $u = q/h$ and the Péclet number
$$\text{Pe} = \frac{UL}{D} \quad (14)$$
measures the relative strength of advective drainage to diffusive redistribution.

Because the characteristic velocity $U$ arises from the gravitational component of the lubrication flux (Sec. II.A), it increases monotonically with inclination. Consequently,
$$\text{Pe}(\theta) \propto \sin\theta, \quad (15)$$
implying that inclination drives the system across three distinct transport regimes:
- Diffusion-dominated regime ($\text{Pe} \ll 1$): concentration gradients relax efficiently, and transport is primarily controlled by confinement through $h(x, t)$.
- Intermediate regime ($\text{Pe} \sim 1$): advection and diffusion compete, generating longitudinal concentration gradients and coupling transport to film morphology.
- Advection-dominated regime ($\text{Pe} \gg 1$): solutes are rapidly displaced downslope, residence times shorten, and the system becomes sensitive to even small variations in inclination.

These regimes arise directly from the nondimensional structure of the governing equations and reflect the geometric amplification of transport associated with increasing $\theta$.

### III.D. Reactive Confinement and the Damköhler Number

Reactive processes within the film introduce an additional nondimensional parameter, the Damköhler number, which compares the timescale of reaction to the timescale of advective transport. For a reaction rate $k$, the nondimensional form is
$$\text{Da} = \frac{kL}{U}. \quad (16)$$
Because the drainage velocity $U$ is set by gravitational forcing and increases monotonically with the inclination angle $\theta$, the Damköhler number decreases accordingly:
$$\text{Da}(\theta) \propto \frac{1}{\sin\theta}. \quad (17)$$

*Contact author: helena.cs.vasconcelos@uac.pt

This inverse dependence has direct implications for reactive confinement:
- Reaction-dominated regime (Da ≫ 1): solute residence times exceed reactive timescales; reactivity is governed primarily by film thickness and intermolecular interactions.
- Transport-limited regime (Da ≪ 1): advection removes solute faster than it can react; reaction becomes constrained by supply rather than kinetics.
- Coupled regime (Da ∼ 1): comparable timescales lead to strong sensitivity to local variations in $h(x,t)$, $u(x,t)$ and confined electrostatics.

The inclination-controlled shift from reaction-dominated to transport-dominated behavior is therefore an intrinsic feature of inclined thin-film systems, arising directly from the nondimensional structure of the governing equations. It also underpins the shortening of reactive intervals and the reduction of wet-state persistence analyzed in Sec. IV.

### III.E. Inclination-driven trends in $\{Bo, Pe, Da\}$ Space

The nondimensional structure derived in Secs. III.A–III.D reveals a coordinated dependence of the Bond, Péclet and Damköhler numbers on the inclination angle. To leading order,
$$Bo \propto \sin\theta, \quad Pe \propto \sin\theta, \quad Da \propto (\sin\theta)^{-1},$$
so that a change in $\theta$ modifies the three dimensionless groups in a fully coupled manner.

These relations follow directly from the nondimensional lubrication flux (Sec. III.A), in which gravitational forcing contributes solely through the term $Bo\sin\theta$, while the velocity scale $U$ inherits the same θ-dependence.

Under the assumptions of the long-wave lubrication approximation (Newtonian viscosity, no slip at the substrate, $h/L \ll 1$, negligible inertia, and source–sink terms that do not introduce additional intrinsic length or velocity scales), this coordinated dependence defines a geometry-imposed, leading-order trajectory in the $\{Da, Pe, Bo\}$ space—shown in Fig. 2.

The shape of this trajectory is universal, in the sense that it follows directly from the structure of the lubrication flux, whereas the position of any specific physical system along it depends on material parameters such as viscosity, intermolecular constants, diffusivity and reaction rates.

Thus, rather than claiming that all thin-film systems evolve identically, we emphasize that inclination constrains the leading-order coupling of Bo, Pe and Da, while material properties determine the quantitative location along the trajectory. The universality arises directly from the structure of the lubrication flux: gravitational forcing grows with inclination (Bo ↑), advection strengthens accordingly (Pe ↑), and the advective enhancement weakens the relative importance of reaction (Da ↓). Together, these coordinated shifts reorganize the balance between drainage, transport and confined reactivity.

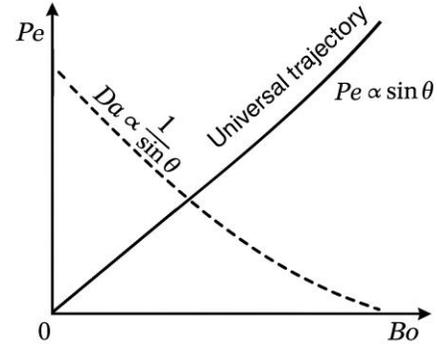

FIG 2. Universal inclination-driven trajectory in the $\{Da, Pe, Bo\}$ parameter space.

The resulting leading-order inclination-driven trajectory provides a compact and physically transparent representation of the underlying dynamics. It captures the progressive transition from capillarity-dominated flow at low inclination to advection-dominated drainage at high inclination, and simultaneously explains the loss of reactive confinement and the shortening of residence times observed in Sec. V. The trajectory therefore offers a unifying perspective on inclined thin-film behavior, linking geometry, force balance and transport hierarchy within a single mathematical structure.

### III.F. Implications for Film Stability, Transport and Rupture

The inclination-driven variations of Bo, Pe and Da summarized by the universal trajectory have direct consequences for film stability, drainage dynamics and confined transport. As $\theta$ increases, gravitational forcing strengthens through Bo, advection becomes progressively dominant through Pe, and the relative importance of reaction diminishes through Da. These coordinated shifts reorganize the hierarchy of physical processes that govern thin-film behaviour.

For stability, an increase in Bo accelerates thinning and promotes earlier access to regimes in which intermolecular attractions dominate. This leads to reduced rupture times and asymmetric dewetting fronts, consistent with the gravitational bias introduced in Sec. II and with theoretical predictions for gravity-driven front propagation and instability asymmetry [4,8].

*Contact author: helena.cs.vasconcelos@uac.pt

For transport, the increase in Pe shortens residence times and enhances the longitudinal advection of solute, weakening the ability of diffusion to maintain uniform concentration profiles. Confined electrostatics and near-surface interactions become less effective as the film is rapidly displaced downslope.

For reactions, the decrease in Da shifts the system from reaction-dominated behaviour at low inclination toward a transport-limited regime at higher inclination. This transition reduces the duration over which solutes can interact within confined, high-resistivity regions and therefore suppresses sustained reactive pathways.

Together, these trends provide a physically cohesive interpretation of the deterministic changes observed in inclined thin-film systems. They also form the foundation for the stochastic wetting analysis developed in Sec. IV, where inclination-driven variations in drainage rate and residence time determine the statistics of wet–dry cycling and the long-time persistence of reactive intervals.

## IV. STOCHASTIC WETTING AND TIME-OF-WETNESS DYNAMICS

Thin films exposed to atmospheric forcing rarely evolve as continuous layers. Instead, they undergo intermittent cycles of formation, drainage, rupture and re-wetting driven by humidity fluctuations, aerosol deposition and environmental noise. While the deterministic equations of Sec. II describe the deterministic dynamics of the film thickness, they cannot capture the discrete transitions between wet and dry states that dominate long-time surface behavior. A stochastic description is therefore required to represent the intermittent nature of atmospheric wetting, particularly on inclined substrates where drainage is strongly enhanced.

Inclination modifies these transitions by reorganizing the drainage flux, accelerating access to rupture thresholds and reducing the residence time of liquid films. The combined deterministic–stochastic structure provides a physically grounded framework for understanding the observed decrease in time-of-wetness with increasing inclination.

### IV.A. Deterministic thin-film dynamics and rupture thresholds

The deterministic dynamics of the film thickness are governed by the lubrication-based thin-film equation
$$\partial_t h + \partial_x q = S(x,t), \qquad (18)$$
which describes the continuous thinning of the film between wetting events. Rupture occurs when the local thickness reaches the critical value $h_{\text{crit}}$, set by the balance between van der Waals attraction and capillary curvature. This threshold marks the limit of metastability for ultrathin films and determines the termination of a wet interval in the absence of re-wetting.

As shown in Secs. II and III, the gravitational contribution to the flux grows with inclination through the term $\text{Bo}\sin\theta$. The increased downslope advection accelerates thinning and causes the deterministic trajectory $h(t)$ to reach $h_{\text{crit}}$ more rapidly as $\theta$ increases. Inclination therefore shortens the deterministic component of the wet-state lifetime, independently of environmental factors such as humidity or aerosol deposition.

The inclination dependence of the drainage time plays a central role in the stochastic framework developed in Sec. IV.B: it sets the scale for the drying rate $\lambda_{\text{dry}}$, which increases monotonically with $\theta$. The deterministic thinning dynamics therefore provide the physical basis upon which the stochastic wet/dry transitions are built.

### IV.B. Stochastic Transition Between Wet and Dry States

Environmental forcing introduces intrinsic randomness into film re-formation events, primarily through humidity fluctuations, aerosol deposition and vapor-phase transport.

To capture this behavior, the surface is idealized as a two-state stochastic process in which the interface alternates between
- a wet state (W), where a continuous or metastable film exists, and
- a dry state (D), where the film has ruptured or is absent.

Transitions between these states occur with rates

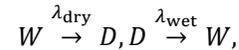

$$W \xrightarrow{\lambda_{\text{dry}}} D, \quad D \xrightarrow{\lambda_{\text{wet}}} W,$$

where $\lambda_{\text{dry}}$ is the drying or rupture rate, and $\lambda_{\text{wet}}$ is the re-wetting rate associated with condensation or deliquescence [19,20].

Inclination modifies these rates asymmetrically. As shown in Secs. II and III, the gravitational contribution to the flux increases monotonically with $\text{Bo}\sin\theta$, reducing the deterministic drainage time of Sec. IV.A and thereby increasing $\lambda_{\text{dry}}$. In contrast, $\lambda_{\text{wet}}$ depends primarily on humidity, temperature and vapor transport, and is only weakly sensitive to $\theta$ [11]. This asymmetry—an inclination-enhanced drying rate together with an approximately inclination-independent wetting rate—is the stochastic counterpart of the inclination-driven re-weighting of the Bond, Péclet and Damköhler numbers established in Sec. III.

*Contact author: helena.cs.vasconcelos@uac.pt

The competition between $\lambda_{\text{dry}}$ and $\lambda_{\text{wet}}$ governs the statistics of intermittent wetting. In particular, the steady-state wet-state probability is

$$P_{\text{wet}} = \frac{\lambda_{\text{wet}}}{\lambda_{\text{wet}} + \lambda_{\text{dry}}}, \quad (19)$$

which decreases systematically with inclination because $\lambda_{\text{dry}}(\theta)$ grows while $\lambda_{\text{wet}}$ remains nearly constant.

This provides a physically grounded explanation for the pronounced reduction in wet-state persistence observed experimentally in inclined electrolyte films [12], without invoking any chemistry-specific mechanisms.

The structure of this two-state stochastic pathway is summarized in Fig. 3. Panel (a) shows the Markov transition scheme, while panel (b) highlights the inclination-induced imbalance between drying and re-wetting: $\lambda_{\text{dry}}(\theta)$ increases with gravitationally driven drainage, whereas $\lambda_{\text{wet}}$ remains largely controlled by humidity. This asymmetry underlies the systematic decline of the wet-state probability with inclination.

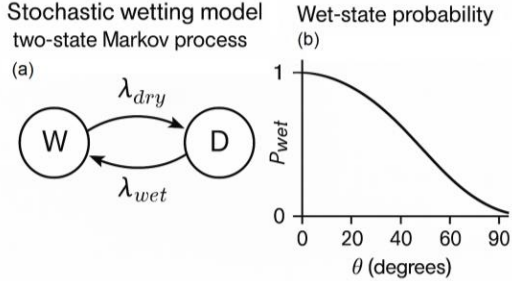

FIG 3. (a) Stochastic wetting model represented as a two-state Markov process, with $\lambda_{\text{dry}}$ denoting the inclination-enhanced drying/rupture rate and $\lambda_{\text{wet}}$ the approximately inclination-independent re-wetting rate associated with condensation;(b) Qualitative dependence of the steady-state wet-state probability $P_{\text{wet}} = \lambda_{\text{wet}}/(\lambda_{\text{wet}} + \lambda_{\text{dry}})$ on inclination angle $\theta$. The monotonic decrease of $P_{\text{wet}}$ reflects the increase of $\lambda_{\text{dry}}(\theta)$ predicted by the thin-film hydrodynamics of Secs. II–III.

Stochastic formulation thus complements the deterministic dynamics by providing a quantitative description of how inclination reshapes the balance between rupture and re-formation events, establishing the foundation for the time-of-wetness analysis in Sec. IV.C.

This two-state model is intentionally minimal and assumes Poissonian (memoryless) transitions, with constant wetting and drying rates. In this formulation, the $W \leftrightarrow D$ dynamics constitute a Markov process with exponential waiting times, providing a mean-field description that neglects temporal correlations and long-tailed statistics. Such a closure is appropriate for capturing inclination-controlled shifts in the average wet persistence, but it does not resolve potential non-Markovian effects, such as heavy-tailed waiting times or humidity-driven correlations associated with environmental fluctuations or substrate heterogeneity. These extensions lie beyond the scope of the present work.

### IV.C. Wet-state persistence and time-of-wetness

The stochastic description introduced in Sec. IV.B determines the persistence of wet states through the competition between the drying and re-wetting pathways. In the long-time limit, the wet-state probability is fully controlled by the ratio $\lambda_{\text{wet}}/\lambda_{\text{dry}}$. Since inclination affects only the drying pathway—by shortening the deterministic drainage time of Sec. IV.A—the wet-state probability inherits a strict geometric dependence, decreasing monotonically with $\theta$.

The predicted variation of the wet-state probability with inclination is shown in Fig. 4.

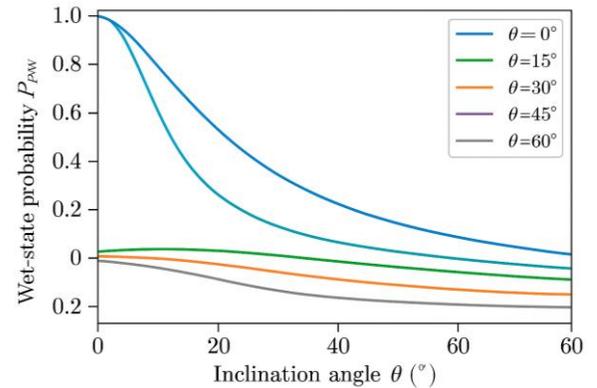

FIG. 4. Steady-state wet-state probability $P_{\text{wet}}$ as a function of inclination angle $\theta$ for several drying/wetting rate ratios. Inclination enhances the drying pathway while leaving the wetting pathway nearly independent of $\theta$, producing a monotonic decline in wet-state persistence.

*Contact author: helena.cs.vasconcelos@uac.pt

The time-of-wetness (TOW), defined as the fraction of a long interval during which the surface remains wet, coincides with this steady-state wet probability for any stationary two-rate process. Thus, TOW decreases systematically with inclination, reflecting the geometry-driven amplification of the drying pathway rather than any chemistry-specific mechanism. This behavior is consistent with the pronounced decline in wet-state persistence observed experimentally for electrolyte films on inclined metallic substrates [11,12].

This probabilistic viewpoint completes the connection between deterministic drainage, stochastic rupture and macroscopic wet/dry statistics, providing the basis for the stability and transport consequences examined in Sec. V.

### IV.D. Coupling Deterministic Dynamics to Stochastic Transitions

The stochastic transitions characterized by the rates $\lambda_{\text{dry}}$ and $\lambda_{\text{wet}}$ acquire physical meaning only when linked to the deterministic thinning behavior established in Sec. IV.A. The drainage time $t_{\text{drain}}(\theta)$—defined as the time for a continuous film to reach its rupture threshold—sets the natural scale for transitions out of the wet state. Consequently,

$$\lambda_{\text{dry}} \sim t_{\text{drain}}^{-1}, \quad (20)$$

so that inclination modifies the drying pathway entirely through its influence on the deterministic fluxes derived in Sec. II and the dimensionless re-weighting discussed in Sec. III.

In contrast, the re-wetting rate is set by vapor-phase processes, condensation kinetics and ambient humidity fluctuations [19,20], none of which couple directly to gravitational asymmetry. The stochastic model therefore inherits a unidirectional sensitivity: inclination accelerates drying but leaves re-wetting largely unchanged.

This separation of roles is essential. It provides a mechanistic interpretation for why wet/dry cycling becomes increasingly intermittent with increasing $\theta$, and why time-of-wetness decreases despite unchanged environmental forcing. The deterministic–stochastic link also ensures that statistical quantities such as the wet-state probability, wet-interval distribution, and the long-time TOW (Sec. IV.C) are governed by geometric rather than material parameters.

### IV.E. Implications for Interfacial Reactivity and Transport

The intermittent wetting generated by the dynamics of Secs. IV.A–IV.D has direct consequences for transport and reactive processes within ultrathin films. Because solute retention, electrostatic confinement and reaction kinetics all depend on the duration and continuity of the wet state, the progressive decline in wet-state persistence with inclination produces a systematic reorganization of interfacial behavior.

*Contact author: helena.cs.vasconcelos@uac.pt

In the low-inclination regime, long and relatively infrequent drying events allow ionic confinement and concentration gradients to develop, favoring reaction pathways that require sustained film continuity. As inclination increases, the shortening of wet intervals limits solute residence time and weakens electrostatic confinement, shifting reactive behavior toward transport-controlled regimes. These effects arise directly from the inclination-controlled imbalance between $\lambda_{\text{dry}}$ and $\lambda_{\text{wet}}$, and not from any material-specific chemistry. This interpretation aligns with experimental observations of reduced wet-state longevity and intermittent reactivity on inclined metallic substrates [11,12].

Stochastic intermittency therefore represents the final element in the multiscale picture developed in this section: deterministic thinning sets rupture timing, stochastic forcing governs re-wetting, and inclination biases the entire process toward rapid cycling. Together, these effects establish the foundation for the stability, drainage and transport consequences analyzed in Sec. V.

More detailed stochastic structure—such as long-tailed wetting intervals, state-dependent rates, or humidity-correlated transitions—falls outside the scope of the present mean-field description and constitutes a natural extension for future work.

### V.A. Stability Regimes Under Inclination

The stability of a thin film on an inclined substrate results from the competition between capillary curvature, intermolecular interactions and the gravitational asymmetry introduced by $\theta$. Linearising the hydrodynamic flux of Sec. II about a uniform base state of thickness $H$ yields the dispersion relation

$$\sigma(k) = -k^4 + \Pi'(H) k^2 + (\text{Bo} \sin\theta) k, \quad (21)$$

where $\sigma(k)$ is the growth rate of a Fourier mode of wavenumber $k$, and $\Pi'(H)$ denotes the derivative of the disjoining pressure evaluated at the base thickness. Capillarity contributes to the stabilizing term $-k^4$, intermolecular forces contribute the destabilizing term proportional to $k^2$, and inclination introduces a linear term that biases the instability toward downhill-propagating modes. The combined effect reshapes both the instability threshold and the spatial structure of the most unstable modes.

For horizontal substrates ($\theta = 0$), the spectrum is symmetric in $k$, and stability is determined solely by the balance between capillarity and intermolecular interactions, recovering the classical result for thin-film rupture [7,9]. This baseline behavior corresponds to the standard long-wave instability structure examined in lubrication-type analyses [8]. Inclination breaks this symmetry: the additional term (Bo $\sin\theta$)$k$ shifts the spectrum and lowers the effective rupture threshold. As a result, perturbations with positive $k$ (downslope) experience enhanced growth, while upslope modes are suppressed. This spectral asymmetry provides a mechanistic explanation for the directionality observed in experiments on inclined wetting films [12] and coating flows [4,5].

The inclination dependence of the maximum growth rate follows directly from the dispersion relation. As $\theta$ increases, the unstable band broadens and its peak shifts to larger $k$, signalling a transition from symmetric dewetting patterns to travelling, downhill-propagating disturbances. The resulting directed rupture fronts are a hallmark of thin films subject to gravitational symmetry breaking. The behaviour predicted by the dispersion relation is illustrated in Fig. 5, which shows the systematic displacement of the instability spectrum with increasing inclination.

This inclination-induced restructuring of the instability landscape establishes the foundation for the drainage, transport and reactive consequences developed in Secs. V.B–V.E.

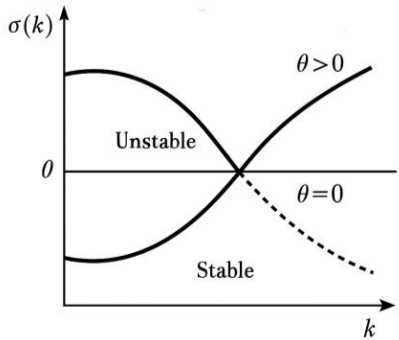

FIG.5. Linear stability spectra $\sigma(k)$ for increasing inclination angles.

### V.B. Drainage and thinning dynamics

Inclination alters the drainage behavior of thin films in a systematic and theoretically transparent manner. Because the gravitationally induced flux term Bo $\sin\theta$ enters the hydrodynamic equation linearly, its influence propagates directly into the thinning kinetics of the film. This linear dependence is the same one that governs the inclination-driven coupling of the Bond, Péclet and Damköhler numbers in Sec. III.E, so the drainage pathway reflects the underlying structure of the nondimensional trajectory.

For a laterally uniform film, the leading-order drainage law obtained from the lubrication equation reduces to
$$\partial_t h \approx -C h^3 \text{ Bo } \sin\theta, \quad (22)$$
where $C$ collects viscosity and geometric constants. This expression encapsulates three key consequences:

**1. Accelerated thinning with inclination**
Since the gravitational contribution scales as $\sin\theta$, the drainage time obeys
$$t_{\text{drain}}(\theta) \propto (\sin\theta)^{-1}. \quad (23)$$
Thus, even moderate inclination substantially shortens the film lifetime, a tendency consistent with the deterministic pathways leading to rupture discussed in Sec. V.A.

This inclination dependence is the drainage counterpart of the universal scaling $Pe \propto \sin\theta$ appearing in the nondimensional formulation.

**2. Enhanced sensitivity in the ultrathin regime**
As the film approaches nanometric thicknesses, the cubic dependence $h^3$ causes the dynamics to stiffen small reductions in $h$ produce disproportionately large increases in thinning rate.

Consequently, inclination-induced acceleration is magnified near the disjoining-pressure–dominated regime, making rupture markedly more abrupt than in horizontal geometries [2].

**3. Earlier loss of continuity and reduced recovery intervals**
Because evaporation competes with downhill drainage, the accelerated thinning induced by inclination shifts the balance strongly toward rupture. Even before stochastic re-wetting is considered (Sec. IV), the deterministic trajectory reaches the rupture threshold $h_{\text{crit}}$ more rapidly.

The subsequent re-formed films originate from progressively thinner initial states, reinforcing the cycle of short-lived wet intervals.

Together, these drainage characteristics supply the deterministic backbone that feeds into the stochastic transitions of Sec. IV.

The inclination-controlled reduction of $t_{\text{drain}}$ is therefore not a system-specific effect but the direct physical manifestation of the universal {Da, Pe, Bo} trajectory derived in Sec. III.

### V.C. Transport regimes: advection, diffusion and confinement

The dimensionless analysis of Sec. III showed that inclination increases the Péclet number through the scaling Pe $\propto \sin\theta$.

*Contact author: helena.cs.vasconcelos@uac.pt

This single geometric modification reshapes the balance between advection and diffusion within the film across three distinct regimes.
These regimes arise from the competition between downhill advection $u = h^{-2}q$ and transverse diffusion and are further modulated by the thickness-dependent confinement of ionic mobility [2,14].
The resulting geometry-induced transport regimes are summarized in Fig. 6.

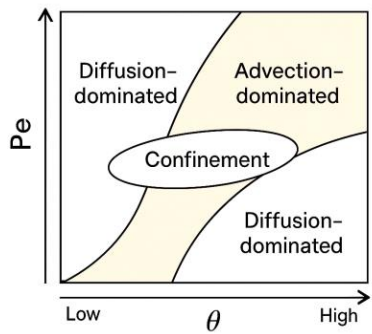

FIG.6. Geometry-induced transport regimes as a function of inclination angle θ and Péclet number Pe.

At low inclination the film is diffusion-dominated; at intermediate inclination mixed advection–diffusion behavior emerges; and at high inclination advection dominates while confinement weakens. The shaded region indicates the thickness range in which ionic mobility remains strongly confined.

### 1. Diffusion-dominated transport at low inclination
For small $\theta$ the advective flux remains weak and
$$\text{Pe} \ll 1.$$
Spatial concentration gradients relax faster than they are generated.
Solutes remain confined inside the film, and their residence time is dictated primarily by the local thickness and molecular diffusivity.
Electrostatic confinement is strong in this regime: the overlap of double layers suppresses mobility and favors transport-limited reactions.

### 2. Mixed advection–diffusion regime at intermediate inclination
As $\theta$ increases, the advective velocity rises proportionally to $\sin \theta$, eventually leading to
$$\text{Pe} \sim \mathcal{O}(1).$$
Diffusion no longer eliminates concentration gradients generated by downhill drainage.
Longitudinal compositional variations appear, solute accumulation zones develop, and reaction rates become coupled to the hydrodynamic timescale. This intermediate regime marks the onset of geometry-induced transport asymmetry: even a nearly uniform film may carry solutes preferentially toward the downhill edge.

*Contact author: helena.cs.vasconcelos@uac.pt

### 3. Advection-dominated transport at high inclination
For large $\theta$,
$$\text{Pe} \gg 1,$$
advection overwhelms diffusion.
Solutes are swept downslope faster than they can redistribute, drastically reducing residence times. Electrostatic confinement weakens because the rapid thinning (Sec. V.B) pushes the film below the stable-confinement regime before diffusive equilibration occurs.
At this point the reaction dynamics are no longer governed by local kinetics but by the rate at which material can be supplied to or retained within the thinning film.

## V.D. Reactive confinement and electrochemical Implications

Although the formulation is general, electrolyte films offer a particularly clear illustration of how inclination reorganizes reactive behavior.
Because the Damköhler number satisfies
$$\text{Da}(\theta) \propto (\sin \theta)^{-1}, \tag{24}$$
increasing inclination reduces the time available for ions to remain confined within the liquid layer. This effect is reinforced by two mechanisms previously established in Secs. IV and V.B: (i) the shortening of wet intervals caused by inclination-enhanced drainage, and (ii) the weakening of electrostatic confinement as the film is driven rapidly into the ultrathin regime.
The combined consequence is a transition from kinetics-dominated behavior at low inclination to transport-limited behavior at high inclination. Reaction pathways requiring sustained connectivity— such as those involving concentrated boundary layers or strongly overlapped double layers—are progressively suppressed. Instead, reactivity becomes governed by the supply of material to the thinning film and by the stochastic occurrence of short wet intervals. These effects do not rely on electrolyte-specific chemistry: they arise generically from the re-weighting of Da, Pe and the rupture rate with $\theta$.
This inclination-driven shift in confinement and reactivity provides a physical interpretation for the reduced wet-state persistence and intermittent electrochemical activity reported experimentally on inclined metallic surfaces [11,12].

## V.E. Intermittency, wet-state persistence and global behavior

The combined deterministic and stochastic descriptions developed in Secs. II–IV reveal a coherent and geometry-driven reorganization of thin-film behavior under inclination. As the Bond and Péclet numbers increase with θ, deterministic thinning accelerates, rupture occurs earlier, and the drying pathway strengthens relative to the re-wetting pathway. The stochastic model of Sec. IV shows that this imbalance raises the drying rate $\lambda_{dry}(\theta)$ while leaving $\lambda_{wet}$ essentially controlled by humidity and vapour transport. The consequence is a systematic decline of the wet-state probability and a transition to highly intermittent wet/dry dynamics.

In this intermittent regime, the long-time behavior of the film is governed by sporadic and short-lived wet intervals separated by extended dry periods. The hydrodynamic structure of the film becomes increasingly asymmetric: downhill thinning accelerates with Pe, confinement breaks down earlier, and the residual film thickness approaches the rupture threshold more rapidly. Transport processes reflect the same geometric influence. At modest inclination, mixed advection–diffusion behavior persists, but at large θ the dynamics become decisively advection-dominated, with solutes swept downslope before diffusive redistribution can occur.

These inclination-induced changes have broad implications for reactive and confined films. Intermittency truncates residence times, interrupts extended reactive phases, and suppresses steady confined-electrostatic regimes. The interface becomes reactive only during brief wet windows, and the corresponding reaction pathways become strongly history-dependent. This global behavior—accelerated thinning, diminished wet-state persistence, loss of confinement, and the emergence of strong intermittency—is a generic consequence of the inclination-driven hierarchy of dimensionless groups, not of any particular chemistry.

## VI. DISCUSSION

The framework developed in this work unifies the hydrodynamic, interfacial, gravitational and stochastic mechanisms governing thin films on inclined substrates. By deriving the governing equations from first principles and analyzing their dimensionless structure, we have shown that inclination acts not as a geometric detail but as a symmetry-breaking field that modifies drainage, stability and transport in a coordinated manner. This reorganization is encoded in a universal inclination-driven trajectory in the {Bo, Pe, Da} parameter space, which emerges directly from the structure of the lubrication-type equations and is therefore independent of system-specific material properties.

This geometric control manifests across all levels of the dynamics. Deterministically, increased inclination enhances gravitational forcing, accelerates thinning, shifts instability thresholds and shortens the residence time of solutes in confined ultrathin regions. Stochastically, the strengthened drainage pathway increases the drying rate while leaving the re-wetting pathway largely unaffected, thereby reducing the steady wet-state probability and enhancing the intermittency of wet/dry cycling. Together, these deterministic and stochastic consequences generate the characteristic features of inclined films: rapid thinning events, asymmetric profiles, premature loss of confinement and fragmented reactive phases.

Representative experimental systems reported in [11,12] fall within the same region of the inclination-driven trajectory in {Bo, Pe, Da} space, reinforcing the physical relevance of the leading-order scaling.

Although the formulation is general, its qualitative implications are consistent with the physical conditions reported for atmospheric electrolyte films. Representative orders of magnitude—typical film thicknesses of $H \approx 10$–$100$ nm, lateral length scales of $L \approx 1$–$10$ mm, and gravitationally driven drainage velocities in the range $10^{-6}$–$10^{-4}$ m s$^{-1}$, all compatible with the regimes discussed in Refs. [11,12]—place such systems in the low-Bond-number range (Bo $\approx 10^{-6}$–$10^{-4}$), with moderate-to-high Péclet numbers (Pe $\approx 1$–$100$) and order-unity Damköhler numbers. These values are broadly aligned with the inclination-dependent trends identified in Sec. III, supporting the physical relevance of the framework while remaining within its intended role as a leading-order, conceptual description rather than a detailed quantitative prediction. These values support the physical relevance of the framework while remaining within the intended conceptual scope rather than attempting detailed quantitative prediction. The stochastic component of the model is deliberately minimal: the wet-to-dry and dry-to-wet transitions are treated as a two-state Poisson process with constant rates. This Markov assumption provides a transparent mean-field description of intermittent wetting, but it does not attempt to capture heavy-tailed waiting-time distributions or humidity-driven temporal correlations. Such effects—often observed in environmental systems—could be incorporated in future work through renewal processes or continuous-time random-walk formulations.

The unified structure presented here also identifies several directions for further development. Extending the analysis to two- and three-dimensional geometries would enable the study of transverse instabilities, rivulet formation and interactions between gravitational forcing and contact-line pinning. Introducing substrate heterogeneity or patterned wettability could reveal non-trivial couplings between microscopic pinning and macroscopic symmetry breaking. Coupling the present framework to detailed electrochemical kinetic models would clarify how intermittent wetting shapes reactive pathways, especially under cyclic humidity forcing. Finally, integrating realistic stochastic humidity profiles may reproduce long-tailed wet-interval distributions observed in environmental systems.

*Contact author: helena.cs.vasconcelos@uac.pt

The present work provides a coherent physical description of thin films under geometric asymmetry, bridging hydrodynamics, intermolecular forces, confined transport and stochastic wetting within a single mathematical structure. The universal inclination-driven trajectory offers a compact conceptual lens through which to interpret a broad spectrum of interfacial phenomena across soft matter, microfluidics, coating flows and atmospheric environments.

## VII. CONCLUSIONS

We have developed a unified theoretical framework that couples hydrodynamics, intermolecular forces, gravitational asymmetry and stochastic wetting to describe the behavior of thin films on inclined substrates. The formulation, derived from first principles, identifies the key dimensionless groups governing the system and shows how inclination systematically reshapes the balance between capillarity, van der Waals forces, drainage and transport.

Inclination enhances the drying pathway without significantly affecting re-wetting, leading to a monotonic reduction of wet-state persistence and the emergence of strongly intermittent wet/dry dynamics. These effects arise directly from the structure of the lubrication-type equations and the associated scalings of the Bond, Péclet and Damköhler numbers.

Although motivated by electrolyte films, the framework is general and applies broadly to confined, evaporating, deliquescent and coating films. By integrating deterministic and stochastic mechanisms within a single structure, this work provides a coherent basis for interpreting thin-film behavior under geometric asymmetry and defines clear directions for future extensions, including three-dimensional geometries, heterogeneous substrates and the coupling to fully resolved reactive models under intermittent wetting.

The aim of this work is to establish a unified and physically transparent theoretical structure rather than to provide fully resolved analytical or numerical solutions. The governing equations derived here enable such analyses—whether asymptotic treatments of specific drainage regimes or full numerical simulations—but these fall outside the scope of the present conceptual, system-independent framework. Future work may exploit the same formulation to obtain explicit solutions for targeted regimes or to quantify inclination effects in specific material systems.

*Contact author: helena.cs.vasconcelos@uac.pt

*Contact author: helena.cs.vasconcelos@uac.pt